\documentclass[smallextended]{svjour3}       % onecolumn (second format)
\smartqed  % flush right qed marks, e.g. at end of proof
\usepackage{graphicx}
\usepackage{epsfig}
\begin{document}

\title{Influence of the Earth on the background and the sensitivity of the GRM and ECLAIRs instruments aboard the Chinese-French mission SVOM}

\author{ Donghua Zhao             \and
        Bertrand Cordier         \and
        Patrick Sizun            \and
        Bobing Wu                \and
        Yongwei Dong                \and
        St\'{e}phane Schanne        \and
        Liming Song                 \and
        Jiangtao Liu
}

\institute{Donghua Zhao (\email{zhaodh@ihep.ac.cn}) \and Bobing Wu \and Yongwei Dong \and Liming Song \and Jiangtao Liu \at
              Key Laboratory for Particle Astrophysics, Institute of High Energy Physics, Chinese Academy of Sciences, 19B Yuquan Road, Beijing 100049, China\\
           \and
            Bertrand Cordier         \and
            St\'{e}phane Schanne     \at
            CEA Saclay, DSM/Irfu/Service d'Astrophysique, 91191, Gif-sur-Yvette, France\\
            \and
            Patrick Sizun            \at
            CEA Saclay, DSM/Irfu/Service d'Astrophysique, 91191, Gif-sur-Yvette, France\\
            CEA Saclay, DSM/Irfu/S\'{e}di, 91191, Gif-sur-Yvette, France
}

\date{Received: date / Accepted: date}

\maketitle

\begin{abstract}
SVOM (Space-based multi-band astronomical Variable Object Monitor)
is a future Chinese-French satellite mission which is dedicated to
Gamma-Ray Burst (GRB) studies. Its anti-solar pointing strategy makes
the Earth cross the field of view of its payload every orbit. In this paper,
we present the variations of the gamma-ray background of the two high energy
instruments aboard SVOM, the Gamma-Ray Monitor (GRM) and ECLAIRs, as a function of the Earth position.
We conclude with an estimate of the Earth influence on their sensitivity and their GRB detection capability.
%\PACS{}
\end{abstract}

%% main text
\section{Introduction}
\label{sec:intro}
After forty years of studies, we are still far from a full understanding of Gamma-Ray Bursts (GRBs). 
Accurate measurements of GRB parameters, such as their position, redshift, peak-energy and so on, 
are needed to further understand GRBs themselves and their use as astrophysical tools. 
SVOM (Space-based multi-band astronomical Variable Object Monitor) \cite{Paul2011}\cite{Basa2008} is 
a future Chinese-French satellite mission which is dedicated to GRB studies.
The payload aboard SVOM consists of two high-energy instruments, the GRM and ECLAIRs,
 together with two low-energy instruments. GRM is a Gamma Ray Monitor, and ECLAIRs is 
a coded-mask camera for X- and Gamma-rays. Together, they can provide GRB localizations 
and GRB spectral observations from a few keV to a few MeV. This wide spectral range is crucial 
to determine as accurately as possible the GRB peak-energy, which is a key parameter to make use of GRBs 
as “standard candles” and extend the measurement of the cosmological parameters at large redshifts \cite{Paul2011}\cite{Ghirlanda2006}.

SVOM is a Low Earth Orbit (LEO) mission with an altitude of 600 km and an inclination of 30$^\circ$. 
Its instruments point close the anti-solar direction during a large fraction of the orbit in order to permit 
the follow-up observations by large ground-based telescopes to optimize the redshift measurements \cite{Paul2011}\cite{Cordier2008}.
This strategy makes the Earth get in and out of the detector’s Field of the View (thereafter FoV) once per orbit. 
The detectors will not operate when crossing the South Atlantic Anomaly (SAA) for protection from the too high 
particle flux, which results in a decrease of the observation time. These factors will cause the instrumental background 
to change when the Earth gets in and out of the FoV and with time. In this paper, we present a study on how the 
gamma-ray background of GRM and ECLAIRs changes with the Earth position in the FoV. The impact of these changes 
on the sensitivity and GRB detection rate are estimated as well. This background research laid the groundwork for the investigation 
of the GRB trigger strategy of GRM and ECLAIRs. This paper is organized as follows:

\S\ref{sec:instr} In this section, we briefly describe the two high energy
instruments aboard SVOM.

\S\ref{sec:simu} The detector mass-models, the simulation method and the
spectral models of gamma-ray background sources are described in this section.

\S\ref{sec:results} The background simulation results, the computations of the
detector sensitivities and the GRB detection rates are presented in this section.

\S\ref{sec:conc} We conclude with a concise summary and a discussion of some
limitations of our method as well as some further improvements.

\section{Instruments}
\label{sec:instr}
\begin{figure}[!ht]
  \centering
  \includegraphics[scale=0.5]{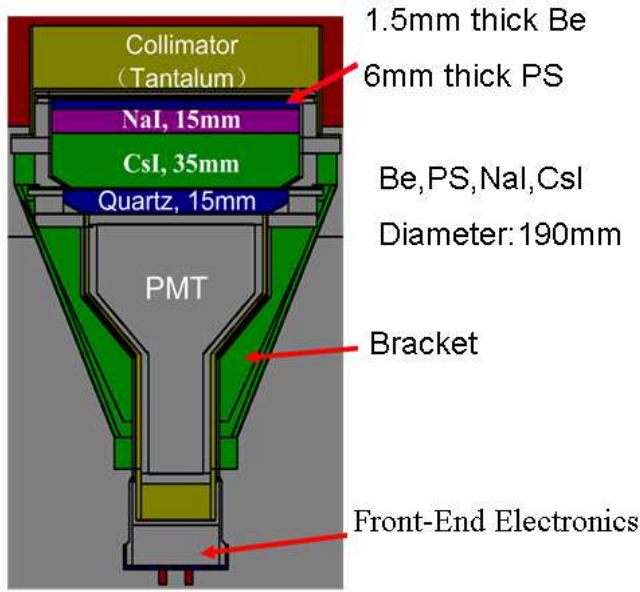}	
  \caption{View of the main components of GRD.}
  \label{fig:Crystal}
 \end{figure}

The main parts of GRM are two identical Gamma Ray Detector (GRD) units, 
which are spectrometers without localization capabilities, sensitive in the energy range 
from ~30 keV to ~5 MeV. Each GRD is equipped with one auxiliary Calibration Detector, 
which mainly consists of a plastic scintillator and an embedded $^{241}$Am source, 
in the collimator above the entrance window. A Charged Particle Detector 
is mounted to alert the high-voltage supplies of PMTs to switch off or on. 
The GRD consists of three scintillator layers (see Figure \ref{fig:Crystal})
which are plastic scintillator, NaI(Tl) and CsI(Na) respectively. They are glued together and viewed with the same light guide coupled to a PMT. 
The pulse shapes generated by the different scintillators can be distinguished by the Pulse Shape Discrimination circuit (PSD). 
The 15 mm thick NaI(Tl) crystal with a diameter of 190 mm works as the main detecting element. 
The CsI(Na) crystal, which is 35 mm thick and has the same diameter as NaI(Tl), 
located behind NaI(Tl) is another important detection element and serves 
also as anti-coincidence element against photons from behind. The element located on top of NaI(Tl) 
is the organic scintillator, dedicated to reject background events due to charged particles. 
The structure of organic scintillator has been modified since the previous design described in \cite{Dong2009} 
cannot shield-off efficiently the photons from the side and its design is more complex. 
The beryllium plate with a thickness of 1.5 mm is chosen as entrance window of this triple phoswich detector. 
A collimator made of tantalum is located in front of the scintillator case to reduce the background by limiting the FoV to 2.5 steradians. 
The main components of GRD are shown in Figure \ref{fig:Crystal}.

ECLAIRs is a coded-mask imaging camera with a 2 sr-wide FoV, with a source localization accuracy better than 10 arcmin, 
sensitive in the 4 keV - 250 keV energy range. Its main entity is an imaging Camera for X- and Gamma-rays (CXG) which 
mainly consists of a detection plane of 80$\times$80 CdTe semiconductor detectors, a coded-mask located above the detection plane, 
a multi-layer lateral shield and mechanical structures, etc. ECLAIRs is also equipped with an onboard 
Scientific Trigger Unit (UTS) \cite{Schanne2007} which seeks in real-time for GRBs and determines their localization for follow-up observations. 
For a detailed description of ECLAIRs, see \cite{Godet2009}\cite{Mandrou2008}.

\section{Monte-Carlo simulations of the instruments gamma-ray background}
\label{sec:simu}
In this section, the mass models of GRM and ECLAIRs used to perform our simulations, 
the simulation method as well as the corresponding background models are described.

\subsection{Mass models and simulation method}
\label{sec:models}
\begin{figure}[!ht]
  \centering
  \includegraphics[scale=0.5]{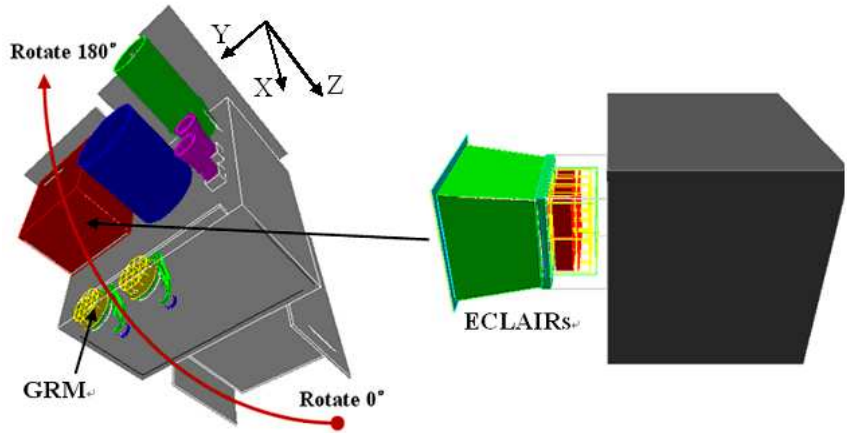}	
  \caption{Schemes of the mass models of GRM and ECLAIRs. (left) Scheme of the
  mass model of GRM (see Section \ref{sec:instr} and Figure
  \ref{fig:Crystal}) as well as the shells of other payloads. The grey boxes correspond to the satellite platform and payload board (not the exact structure of the spacecraft). 
(right) Scheme of the mass model of CXG as well as the cold plate and the readout electronics box placed under the camera. The gray box corresponds to the satellite platform.}
  \label{fig:MassModel}
 \end{figure}

The simulations have been performed with release 4.9.3p02 of the Geant4 toolkit \cite{Agostinelli2003}.
In accordance with the design of GRM, whose components and main characteristics
are described in Section \ref{sec:instr}, its mass model was built. In order to
be close to reality, the other payloads, the payload board and the satellite platform are also considered in the mass model, as shown in Figure \ref{fig:MassModel}
(left). But currently only GRDs are detailed, the others are represented by shells of aluminum with shapes and thickness corresponding to their latest design. 

For ECLAIRs, the CdTe detectors, the ceramics, the ASICs, the multi-layered shield composed of 
Pb/Cu/Honeycomb/Cu, the coded mask made of a Ta-W alloy, the multi-layer thermal coating insulation (MLI) 
above the coded mask, the TA6V upper and lower supports of the mask, as well as simplified versions of 
the cold plate in AlBeMet and the readout electronics boxes are considered in the mass model. 
The platform is represented by a 1 m$^3$ and 300 kg pseudo-aluminum cube.

In the background simulations, the most important point is to study how the background varies with the Earth position in the FoV.
In practice, we keep the satellite static and let the Earth rotate around it. The establishment of the coordinate system is shown in
Figure \ref{fig:MassModel}: the Z axis points to the opposite direction of the FoV and the coordinate origin is at the center of the satellite.
When the Earth rotates from Z to -Z, the angle changes from 0$^\circ$ to 180$^\circ$.
We divide the half orbit equally into 7 pieces, with an angle-interval of 25.7$^\circ$.
For each piece of orbit, the Earth position can be represented by a typical position approximately.
And the zenith angle of each typical position in the coordinate is used to express the Earth position.
The typical Earth position of Rotate 0$^\circ$ means that the Earth is on Z and Rotate 180$^\circ$ is the Earth position which is on -Z.
Other typical positions in between are Rotate 30$^\circ$, Rotate 60$^\circ$, Rotate 90$^\circ$, Rotate 120$^\circ$ and Rotate 150$^\circ$
respectively. When the Earth rotates from -Z to Z, this half orbit is also divided equally into 7 pieces as the same as that from Z to -Z.
Hence the 7 pieces from -Z to Z are axisymmetric with those from Z to -Z respectively, and can still be represented by the 7 typical positions
(Rotate  0$^\circ$ -  Rotate 180$^\circ$). Here we assume that for different Earth positions with the same zenith angle but different
azimuth angles, the background level is the same. With the assumptions and simplifications above, we can now simulate the background levels
with the 7 typical Earth positions, and each position lasts 1/7 of the orbital period.
These assumptions will be used in Section \ref{sec:sensitivity} to derive a first order estimation of GRB detection rate.

\subsection{Space environment}

The satellite will be subject to different sources of background in space: (i) extragalactic components including 
cosmic X- and gamma-rays, protons, electrons, positrons, and cosmic He-nuclei, etc.; (ii) near Earth components 
consisting of albedo gamma-rays, albedo neutrons, albedo electrons, albedo positrons and the secondary protons 
located under the radiation belt; (iii) as well as the particles from the Sun. 

In this paper three kinds of gamma-ray background are investigated in detail, which are the main background 
components in the enregy range of our interest and include cosmic gamma-rays, albedo gamma-rays and reflected cosmic gamma-rays. 
We adopt the latest spectral models which are Moretti2009 \cite{Moretti2009}, Sazonov2007 \cite{Sazonov2007} and Churazov2006 \cite{Churazov2006}\cite{Churazov2007} respectively. 
For convenience, we will denote cosmic gamma-ray background, reflected cosmic gamma-ray background and albedo gamma-ray 
background as CXB, REFLECTION and ALBEDO respectively in the following context, and if there is no specific declaration the term background 
only means the gamma-ray background.

For ECLAIRs, the background is dominated by the gamma-ray background below 100 keV outside SAA. And the proportion of the background due to
the direct interactions with particles such as electrons, protons and neutrons is about 10\% \cite{Godet2009}\cite{Godet2005}. Concerning the delayed background mainly
introduced by the trapped particles inside SAA, it is outside the scope of this work and we will not discuss it in detail. 
For GRM, in addtion to our simulations of gamma-ray background, G. Li estimated the background induced by the cosmic protons and electrons, 
the secondary protons and electrons, as well as the trapped protons in SAA, when the Earth is at Rotate 0$^\circ$ 
using the method provided in \cite{Lg2009}. With these two parts of work, we obtained that (G. Li, personal communications, 2011)
the average proportion of the gamma-ray background in all kinds of background is approximately 80\% in the energy range 50 keV-300 keV for GRM\_NaI. 
The average proportion of the delayed background caused by the trapped protons in SAA region is less than 15\% below 300 keV, 
and is approximately 50\% above 550 keV. For GRM\_CsI, the proportion of the delayed background is slightly higher.
The background which is used to do calculations in Section \ref{sec:sensitivity} only includes the three kinds of gamma-ray background.

\subsubsection{Cosmic Gamma-ray Background model}

The spectral model of Moretti2009 which is described by Equation (1) is adopted for CXB simulations. 
This model was produced by taking advantage of both Swift-XRT and Swift-BAT databases in the energy band 1.5-200 keV. 
The spatial distribution is isotropic in a 4$\pi$ solid angle. The Earth occultation is considered in the post-processing after the simulations with Geant4.

\begin{equation}
 \frac{dN}{dE}=\frac{0.109}{(\frac{E}{29.0})^{1.40} + (\frac{E}{29.0})^{2.88}}	[ph/keV/cm^2/s/sr]
\end{equation}

Ajello2008 \cite{Ajello2008} is another recent model with a smoothly joined double power law.
It was obtained by fitting the data in the 2 keV-2 MeV range. Below 500 keV, the relative error is less than 20\% 
using model Moretti2009 and Ajello2008. And above 500 keV, the statistical fluctuation is much larger than the relative error. 
Other models such as Sreekumar1998 \cite{Sreekumar1998} and Gruber1999 \cite{Gruber1999} were also compared.
Since there is not much difference using these models for GRM, 
and considering that Moretti2009 is a credible model for ECLAIRs, 
we finally determined to use the model Moretti2009 for both GRM and ECLAIRs.

\subsubsection{Reflected cosmic gamma-ray model}

The outer layers of the Earth's atmosphere reflect part of the incident X- and gamma-ray spectrum via Compton scattering. 
The spectrum in the 1-1000 keV energy range reflected by the Earth's atmosphere was calculated by Churazov et al. (2006). 
The ratio of the reflection spectrum to the incident spectrum is approximated by Equation (2). The incident spectrum is the CXB model suggested 
by Gruber et al. (1999). The reflection is most significant near 60 keV and declines towards lower or higher energies. 
Since the incident directions of photons are connected with Earth positions,
some methods used in simulations are similar to those for Albedo photons
described in Section \ref{sec:albedo}.

\begin{equation}
 A(E)=\frac{1.22}{ (\frac{E}{28.5})^{-2.54} + (\frac{E}{51.3})^{1.57} - 0.37} \times \frac{ 2.93 + (\frac{E}{3.08})^{4} }{1 + (\frac{E}{3.08})^4 } \times \frac{0.123 + (\frac{E}{91.83})^{3.44} }{ 1 + (\frac{E}{91.83})^{3.44}}
\end{equation}

\subsubsection{Albedo gamma-ray model}
\label{sec:albedo}
The albedo gamma-ray background consists of the atmospheric hard X-ray emission 
which is induced by cosmic rays such as energetic protons, alpha particles, 
cosmic electrons and positrons as well as secondary electrons and positrons.

Sazonov et al. found that the shape of the spectrum emergent from the atmosphere 
in the range 25-300 keV is mainly determined by Compton scattering and photo-absorption, 
and is almost insensitive to the incident cosmic-ray spectrum. They provided a fitting formula 
for the hard X-ray surface brightness of the atmosphere which would be measured 
by a satellite-borne instrument, as a function of energy, solar modulation level $\Phi$ \cite{Usoskin2005}, 
geomagnetic cut-off rigidity R$_{cut}$ and zenith angle $\theta$; see
Equations (\ref{equ:albedo1}), (\ref{equ:albedo2}) and (\ref{equ:albedo3}).

\begin{equation}
\label{equ:albedo1}
\frac{dN}{dE}=\frac{C}{ (\frac{E}{44keV})^{-5} + (\frac{E}{44keV})^{1.4} }  	 [ph/cm^2/s/sr/keV]	  
\end{equation}

\begin{equation}
\label{equ:albedo2}
 C=\frac{3\mu(1+\mu)}{5\pi} \frac{1.47\times 0.0178\times [(\Phi/2.8)^{0.4} + (\Phi/2.8)^{1.5}]^{-1} } {\sqrt{1 + [\frac{R_{cut}} {1.3\Phi^{0.25} (1+ 2.5\Phi^{0.4}) }]^{2}}}  	[ph/cm^2/s/sr]
\end{equation}

where, $\mu = cos\theta$

\begin{equation}
\label{equ:albedo3}
 R_{cut} = 59.4 \times \frac{cos^4\lambda_{m}}{ [1 + (1+ cos^3\lambda_{m} sin\theta sin\xi )^{1/2}]^2 }  	[GV]
\end{equation}

\begin{figure}[!ht]
  \centering
  \includegraphics[scale=0.3]{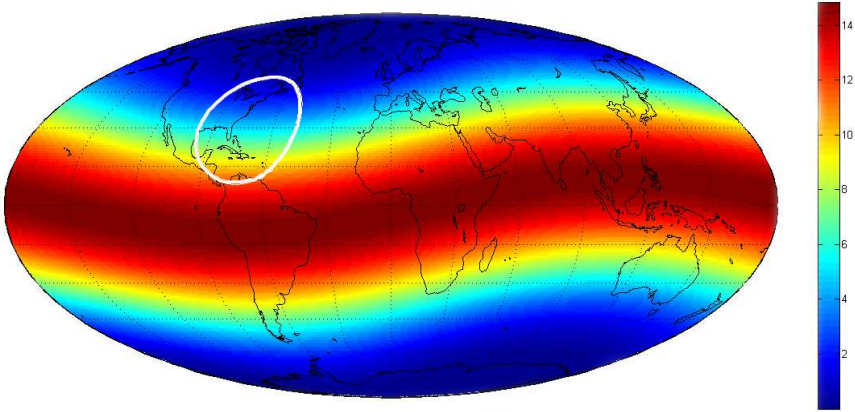}
  \caption{Geomagnetic cut-off rigidity (unit: GV) map. The region surrounded by a white line is the
  atmosphere spherical cap visible by SVOM when the satellite position is 30$^{\circ}$N, 287$^{\circ}$E.}
  \label{fig:Cap}
 \end{figure}

Equation (\ref{equ:albedo2}) allows us to predict the hard X-ray surface brightness
of the atmosphere or its part when observed from a spacecraft, provided that its orbit is higher than 50 to 100 km. 
As a practical example, Sazonov et al. used this formula to predict the hard X-ray flux which would 
be measured by a satellite-borne instrument whose FoV covers the entire terrestrial disc. 
In our case, SVOM is a LEO mission with an altitude of 600 km and an inclination of 30$^\circ$. 
If we suppose that SVOM is aiming towards the nadir direction, only a part of the atmosphere is visible by the instruments. 
Taking into account that mainly the upper part of the atmosphere is producing the hard X-ray emission, 
we consider that the albedo source is at an altitude of 100 km. With a satellite altitude of 600 km the projected radius 
of the visible atmosphere spherical cap is $\sim$2400 km, see the region circled
by white line in Figure \ref{fig:Cap}.

For a given position of the satellite on the orbit, we have to locate the atmosphere spherical cap visible to the instruments. 
Inside this spherical cap, the magnetic latitude $\lambda_m$ varies and therefore the geomagnetic cut-off rigidity is not uniform, as shown in Figure \ref{fig:Cap}. 
For different satellite positions, we do not intercept the same area of the geomagnetic rigidity map. 
As a result, the global ALBEDO level varies with the satellite position along the orbit.

{\bf Hard X-ray surface brightness use}

In order to compute the hard X-ray surface brightness at a given satellite position we propose the following methodology:

For a given satellite position, we define a new reference frame OXYZ in a spherical coordinate system with the 
Z axis going from the Earth center trough the SVOM satellite. The atmosphere elementary elements of the spherical cap 
are represented by the coordinates $\theta_{1}$ and $\varphi_{1}$ in the OXYZ frame (see Figure 4.). 
In practice we divided the atmosphere spherical cap into 2400 elementary areas with 20 $\theta_{1}$ and 120 $\varphi_{1}$ steps 
(a compromise between accuracy and computation time). For each elementary area, we do the calculations as follows:

\begin{itemize}
 \item [-] We determine the geomagnetic latitude $\lambda_{m}$ by performing a reference frame 
change from the OXYZ frame to the geomagnetic reference frame. The reference year for the geomagnetic 
reference frame is 2015 when the Earth's magnetic South Pole is located at (80.367$^\circ$N, 72.624$^\circ$W)\footnote{http://wdc.kugi.kyoto-u.ac.jp/igrf/gggm/}.
 
\item [-] Knowing $\lambda_{m}$, we calculate R$_{cut}$ which depends on the zenith angle $\theta$ 
and on the azimuth angle $\xi$, see Equation (\ref{equ:albedo3}). 
Following the simplification proposed by Sazonov et al. we assume $\xi$ = 0. 

\item [-] We compute the brightness according to Equation (\ref{equ:albedo2}).
And the surface-brightness map corresponding to the satellite position (30$^\circ$N, 287$^{\circ}$E) is represented in Figure \ref{fig:Brightness}.
\end{itemize}

  \begin{figure}[!ht]
  \centering
  \includegraphics[scale=0.3]{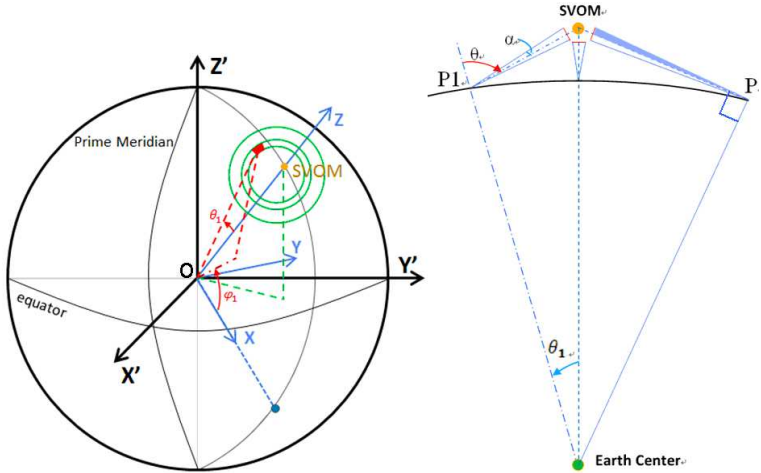}
  \caption{Schematic drawing of albedo gamma-ray background source model. 
(left) Three-dimensional schematic drawing of the albedo gamma-ray background source model. 
The fan-shaped area is the elementary element defined by $\theta_1$ and $\varphi_1$. 
The position of ($\theta_1$, $\varphi_1$=0) corresponds to the point of (satellite latitude - $\theta_1$, satellite longitude). 
(right) Plane schematic drawing of the albedo gamma-ray background source model 
(the straight line around SVOM represents the radiation source in the simulation model built in Geant4. In this work, the line P-SVOM is 
the borderline of the photon directions).}
  \label{fig:sr}
 \end{figure}

  \begin{figure}[!ht]
  \centering
  \includegraphics[scale=0.2]{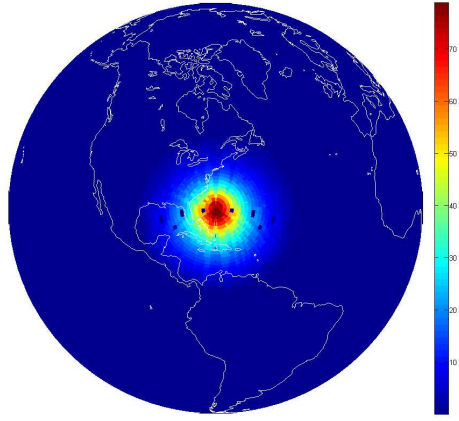}
  \caption{Hard X-ray surface brightness (unit: $10^{-4} ph/cm^2/s/sr$) at the satellite position (30$^\circ$N, 287$^\circ$E).}
  \label{fig:Brightness}
 \end{figure}

\begin{figure}[!ht]
  \centering
  \includegraphics[scale=0.3]{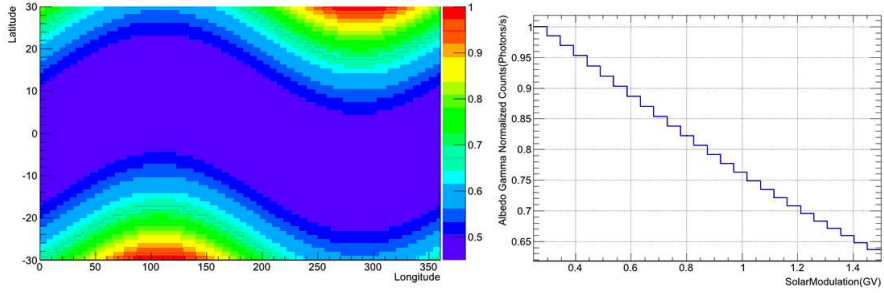}
  \caption{(left) The distribution of the global hard X-ray surface brightness along the satellite orbit. 
The highest level position is (30$^\circ$N, 287$^\circ$E), and the lowest level position is (7$^\circ$N, 151$^\circ$E). 
(right) Variations of the global brightness as a function of the solar modulation when the satellite is at (30$^\circ$N, 287$^\circ$E).}
  \label{fig:Solar}
 \end{figure}

For a given satellite position, by summing the contribution of each cell, we can compute the global hard X-ray surface brightness (ph/s). 
The highest level is obtained at the satellite position (30$^\circ$N, 287$^\circ$E) and the lowest level is obtained at (7$^\circ$N, 151$^\circ$E), 
see Figure \ref{fig:Solar} (left). Moreover, the brightness decreases as the solar modulation $\Phi$ increases from 0.25 GV to 1.5 GV, see Figure \ref{fig:Solar} (right).
In order to study the variation of the ALBEDO as a function of the Earth position, we decided to 
choose the most unfavorable case: we fixed the satellite position at (30$^\circ$N, 287$^\circ$E) with solar modulation 0.25 GV.

{\bf Albedo gamma-rays computation}

For the chosen satellite orbit position, the above computations enabled us to find out the value of all parameters in Equations (\ref{equ:albedo1}-\ref{equ:albedo3}) 
and to compute a surface-brightness map providing us, for each elementary atmospheric cell of center P1 characterized 
by coordinates ($\theta_{1},\varphi_{1}$) and area A($\theta_{1},\varphi_{1}$), with a photon spectrum $\frac{dN}{dE}$($\theta_{1},\varphi_{1}$; E) 
in unit of photons/cm$^2$/s/sr/keV.

Considering that our Geant4 model of the SVOM satellite is enclosed in a sphere, we can compute the half-angle $\alpha$ of the cone 
subtended by the satellite from the center of each cell (see Figure
\ref{fig:sr}, right), and thus its solid angle $\Omega(\theta_{1},\varphi_{1}) = 2\pi(1-\cos\alpha)$.

After choosing an exposure time, with the spectrum $\frac{dN}{dE}$($\theta_{1},\varphi_{1}$; E), the solid angle $\Omega(\theta_{1},\varphi_{1})$ and 
the area A($\theta_{1},\varphi_{1}$) of each elementary atmospheric cell, we can compute the total number of 
photons to shoot and the relative weight of each cell. Finally, we can generate the appropriate number of photons for each cell, 
randomly giving them the appropriate energy, direction and position. Given the chosen satellite attitude - the position of 
the Earth in the FoV, we apply a suitable coordinate transformation before injecting these photons into our Monte-Carlo model.

\section{Results from gamma-ray background simulations}
\label{sec:results}

\begin{figure}[!ht]
  \centering
  \includegraphics[scale=0.5]{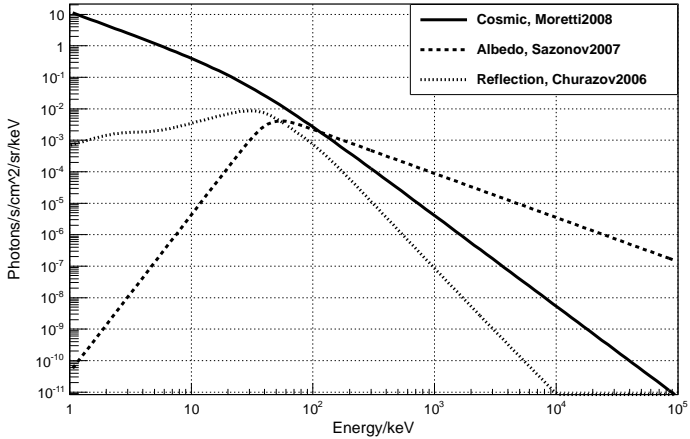}
  \caption{Input energy spectra of the three kinds of gamma-ray background sources.}
  \label{fig:Spectra}
 \end{figure}

Considering the sensitive energy ranges of instruments, the input energy spectra are extended to [10 keV, 100 MeV] for GRM 
and to [1 keV, 100 MeV] for ECLAIRs in the simulations. The input spectra of CXB and REFLECTION were obtained directly from 
their energy spectral equations. For ALBEDO, the spectrum in Figure \ref{fig:Spectra} is the average spectrum from the 
whole atmosphere spherical cap visible by SVOM when the satellite position is (30$^\circ$N, 287$^\circ$E) and solar 
modulation is 0.25 GV. Comparing the three spectra, we can see that the CXB flux is the highest at low energy ($< \sim$100 keV), 
the ALBEDO flux becomes highest as energy increases, and the REFLECTION has an intermediate level at low energy. These features imply 
that the ALBEDO will play a very important role in the high-energy range. The difference between these three kinds of background 
is not only the generation mechanisms but also the spatial distributions. This suggests their influence on the final instrument background 
will be varying with the Earth position. In this section, the instrumental backgrounds of GRM and ECLAIRs induced by these three kinds of 
background sources are described. In Figure \ref{fig:NaI2}, \ref{fig:CsI2} and
Figure \ref{fig:Cxg2}, we only show the results corresponding to Rotate
0$^\circ$, Rotate 90$^\circ$, Rotate 120$^\circ$ and Rotate 180$^\circ$, other Earth positions lead to very similar spectra.
The counting rates of each background component in various energy ranges corresponding to 
the Earth positions of Rotate 0$^\circ$ and Rotate 180$^\circ$ are described in
Table \ref{tab:1}.

\subsection{GRM gamma-ray background}
\label{sec:GRM}
\begin{figure} 
    \centering 
 \includegraphics [scale=0.55, angle=0.0]{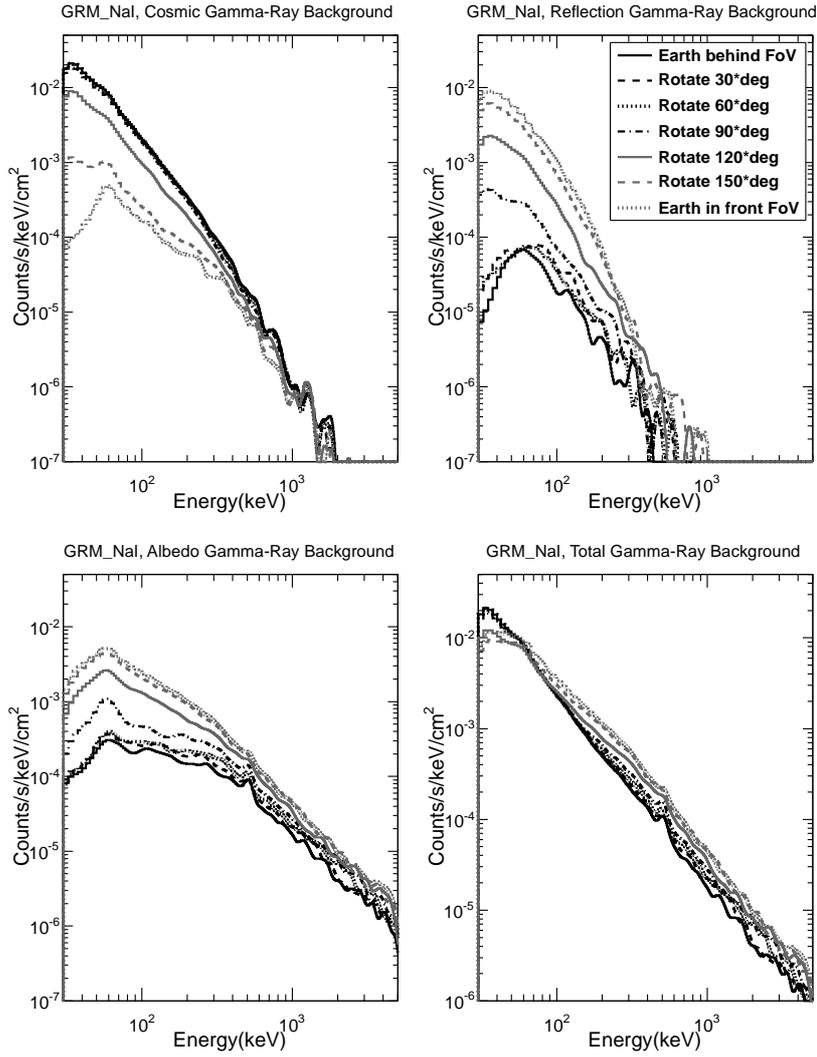}
  \caption{The gamma-ray background levels of GRM\_NaI change with the Earth positions.
(left top) CXB of GRM\_NaI changes with the Earth positions.
(right top) REFLECTION of GRM\_NaI changes with the Earth positions.
(left bottom) ALBEDO of GRM\_NaI changes with the Earth positions.
(right bottom) The total gamma-ray background level of GRM\_NaI changes with the Earth positions.
In Figure \ref{fig:NaI1} - Figure \ref{fig:Cxg2}, the meanings of different lines shown in the legend on the corresponding top right panel are valid for all the four panels.
}
  \label{fig:NaI1}
  \end{figure}

\begin{figure}[!ht]
  \centering
 \includegraphics [scale=0.55]{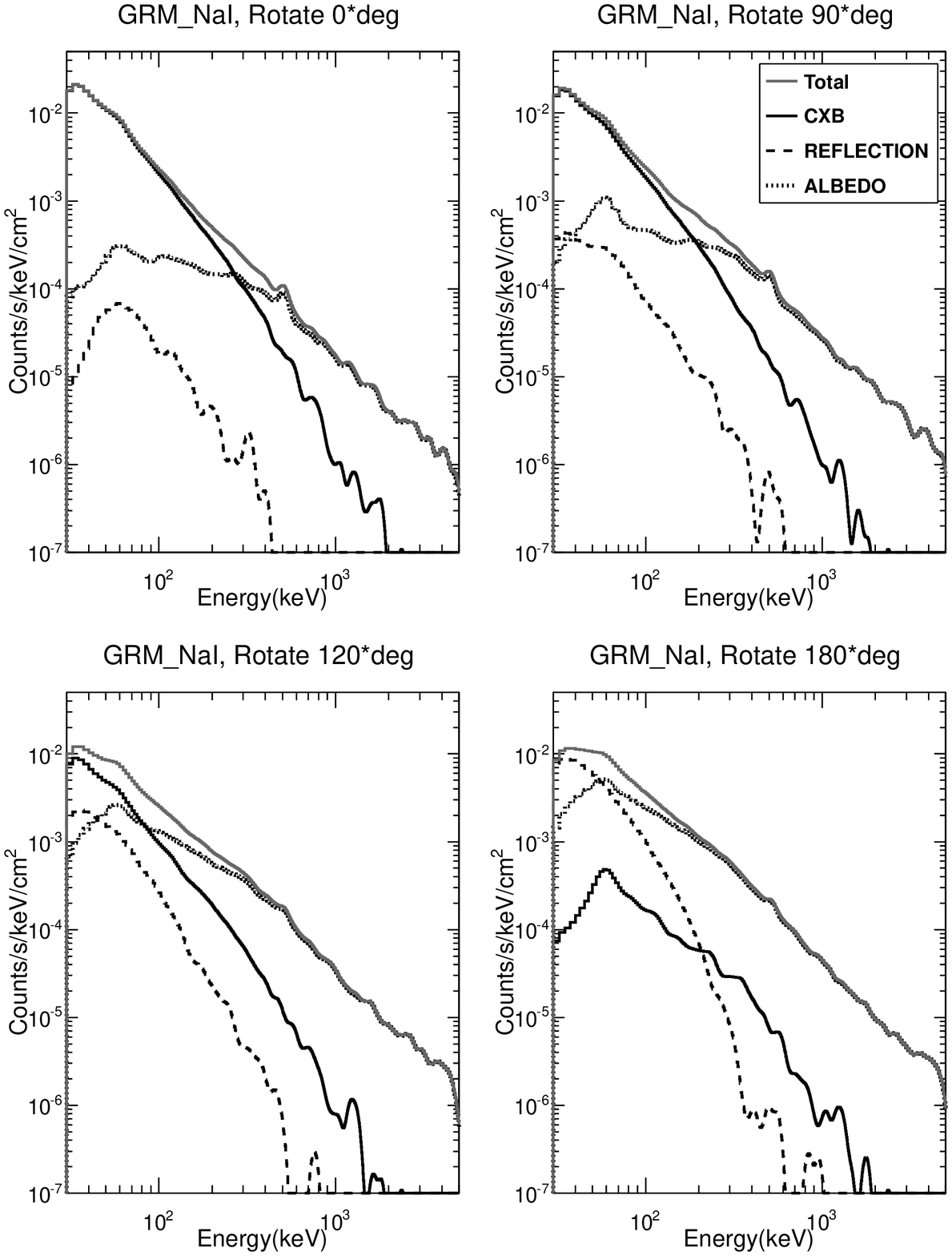}
  \caption{Three kinds of gamma-ray background of GRM\_NaI change with the Earth positions.}
  \label{fig:NaI2}
 \end{figure}

GRM has two detecting elements NaI(Tl) and CsI(Na). Their backgrounds are shown and discussed respectively. 
Figure \ref{fig:NaI1} and Figure \ref{fig:NaI2} present the changes of the three kinds
of background in the NaI as well as the total background with Earth positions.

From Figure \ref{fig:NaI1}, the CXB reaches the maximum level when the Earth is at
Rotate 0$^\circ$ and reaches the minimum when the Earth is at Rotate 180$^\circ$. On the contrary, the ALBEDO and REFLECTION reach their maximum level 
when the Earth is at Rotate 180$^\circ$. Figure \ref{fig:NaI2} shows that the
CXB decreases and the other two backgrounds increase quickly and obviously when the Earth rotates more than 90$^\circ$. Finally when the Earth gets to Rotate 180$^\circ$, 
the CXB becomes the lowest background component at a relatively low energy range (less than 200 keV), 
while the ALBEDO is always the highest at relatively high energies (more than 300 keV). 
That is because its flux is always the highest one at energies above 100 keV.
The REFLECTION, which is always the lowest background component above 200 keV, 
starts to become higher than the ALBEDO below 50 keV when the Earth starts to enter the FoV.

From Figure \ref{fig:NaI1}, for each separate background we can see obvious
differences between Earth positions only at relatively low energies (below 300 keV) instead of the whole energy band. That is because the shields around 
the detector are not as effective for high-energy photons as for low-energy photons. For the same reason we cannot 
see obvious differences in the background when the Earth rotates less than 90$^\circ$ from the back of GRM.

The emission peak of 57 keV in the CXB spectrum is due to the element Tantalum in the collimator. 
When the Earth enters the FoV, less cosmic photons can enter NaI directly through the FoV, 
and relatively the proportion of counts induced by the photons, which are generated by Compton scattering of the relatively high energy photons 
with CsI or other surroundings, becomes larger at low energy. Overall, Earth occultation makes the counts at low energy decrease,
and the photons which interact with Tantalum in the collimator don't vary significantly. Consequently, the proportion of the counts caused by
Tantalum fluorescence becomes higher and the peak becomes more obvious as the Earth enters the FoV.
The 511 keV peak caused by electron-positron annihilation is only obvious 
in the ALBEDO spectra. That is because the photons with energy more than 1022 keV from the ALBEDO source are much 
more numerous than those from the other two background sources. The 511 keV peak is detectable in the total background spectra, 
and can be used for energy calibration.

The total background is shown in Figure \ref{fig:NaI1} (right bottom). At low
energy (less than 50 keV), the background level is highest when the Earth is at Rotate 0$^\circ$. At an energy above 50 keV, 
the highest background is obtained when the Earth is at Rotate 180$^\circ$.

\begin{figure}[!ht]
  \centering
  \includegraphics [scale=0.55, angle= 0.0]{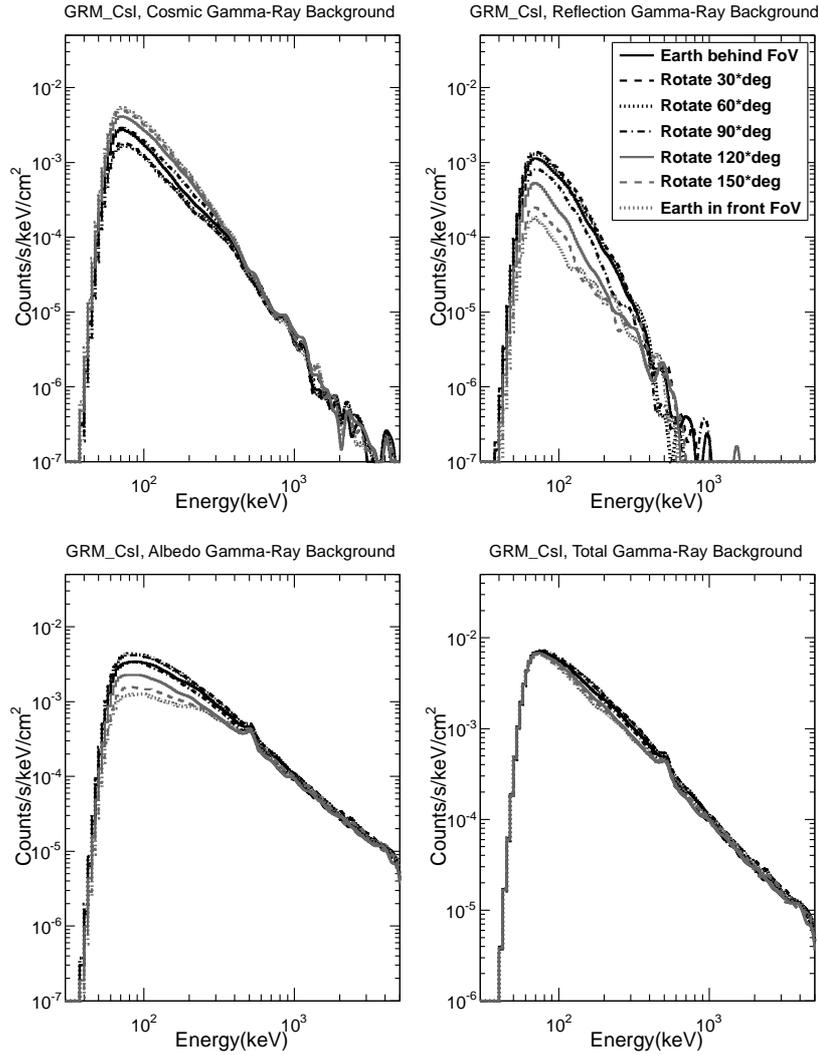}
  \caption{The gamma-ray background levels of GRM\_CsI change with the Earth positions.
(left top) CXB of GRM\_CsI changes with the Earth positions.
(right top) REFLECTION of GRM\_CsI changes with the Earth positions.
(left bottom) ALBEDO of GRM\_CsI changes with the Earth positions.
(right bottom) The total gamma-ray background level of GRM\_CsI changes with the Earth positions.}
  \label{fig:CsI1}
 \end{figure}

\begin{figure}[!ht]
  \centering
  \includegraphics [scale=0.55]{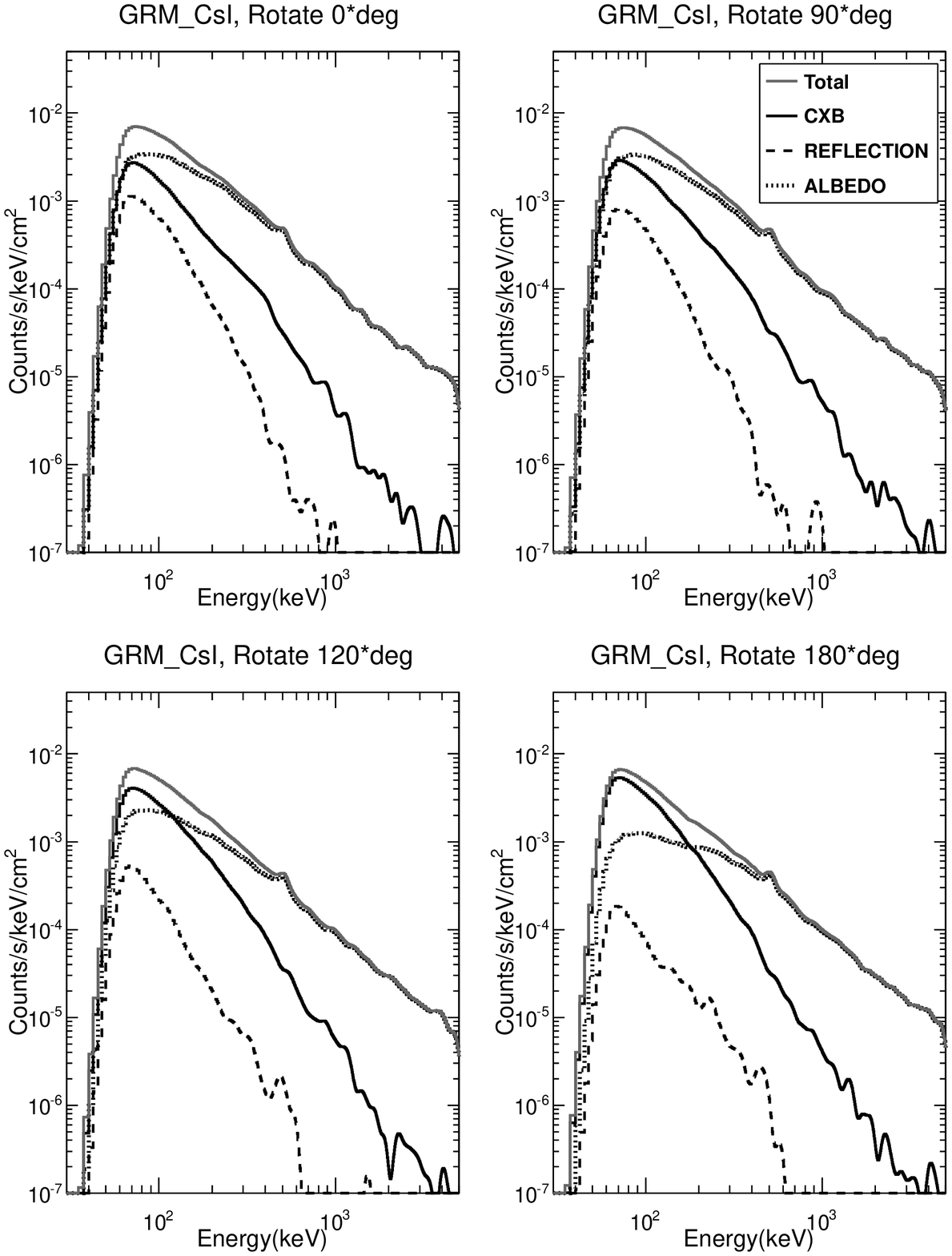}
  \caption{Three kinds of gamma-ray background of GRM\_CsI change with the Earth positions.}
  \label{fig:CsI2}
 \end{figure}

Figure \ref{fig:CsI1} and Figure \ref{fig:CsI2} present how the background of CsI changes with Earth positions. 
In general, the background of CsI is higher when the background source is outside the FoV than that when it is inside the FoV. 
The ALBEDO and REFLECTION components reach their maximum (minimum) levels when the Earth is at Rotate 60$^\circ$ 
(Rotate 180$^\circ$) in 60 keV-250 keV. And in this energy range, the CXB component reaches the maximum (minimum) level when the Earth is at 
Rotate 180$^\circ$ (Rotate 60$^\circ$). That is because of the design of CsI and its position inside GRM (see Figure \ref{fig:Crystal}). 
On one hand, NaI which is in front of CsI can reduce the background of CsI from the FoV, 
and the structures such as the front-end electronics and so on which are behind CsI can decrease the background from behind. 
One the other hand, the interaction cross-section of CsI is the largest when the photons come from the direction of Rotate 60$^\circ$. 
See Figure 11, ALBEDO and REFLECTION decrease as the Earth moves close to the FoV when the CXB increases gradually. 
Finally, the CXB becomes the highest background at relatively low energies (below $\sim$200 keV). 
The ALBEDO is always the highest background at relatively high energies (above $\sim$200 keV). 
The REFLECTION is the lowest component in the whole energy range.
And the total background of CsI is shown in Figure \ref{fig:CsI1} (right bottom).

  \begin{figure}[!ht]
  \centering
  \includegraphics [scale=0.55, angle= 0.0]{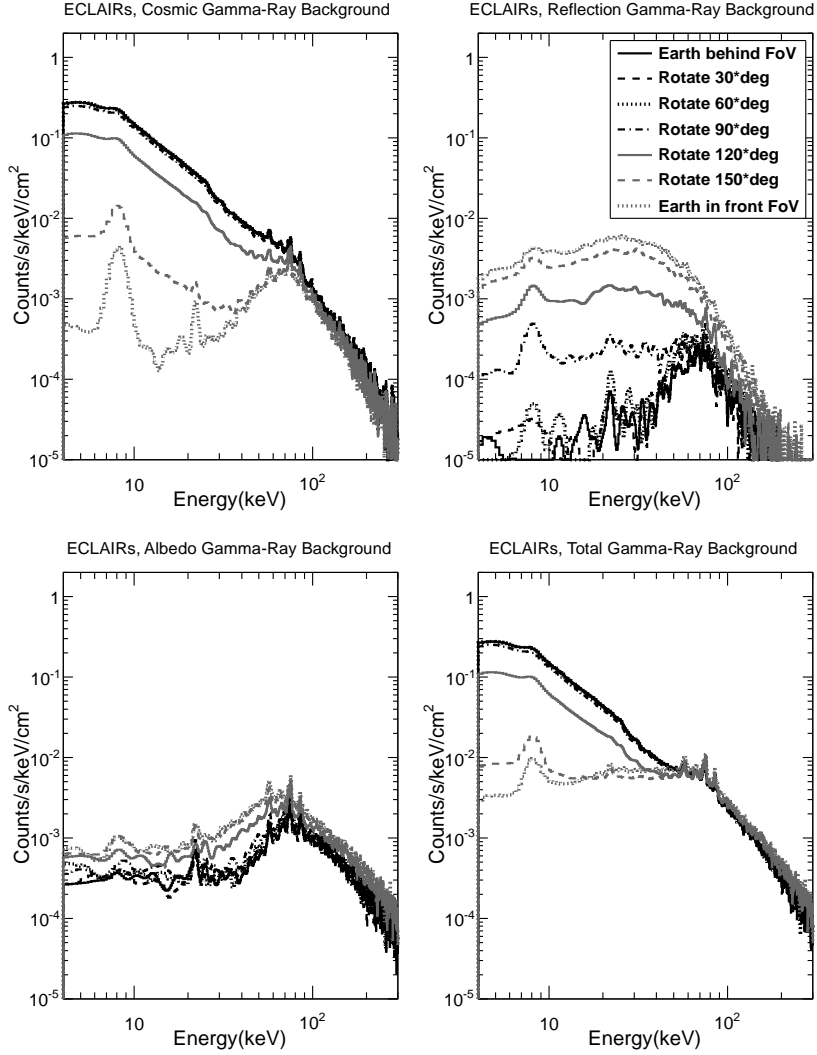}
  \caption{The gamma-ray background levels of ECLAIRs change with the Earth positions
(left top) CXB of ECLAIRs changes with the Earth positions
(right top) REFLECTION of ECLAIRs changes with the Earth positions
(left bottom) ALBEDO of ECLAIRs changes with the Earth positions
(right bottom) The total gamma-ray background level of ECLAIRs changes with the Earth positions.}
  \label{fig:Cxg1}
 \end{figure}

  \begin{figure}[!ht]
  \centering
  \includegraphics [scale=0.55]{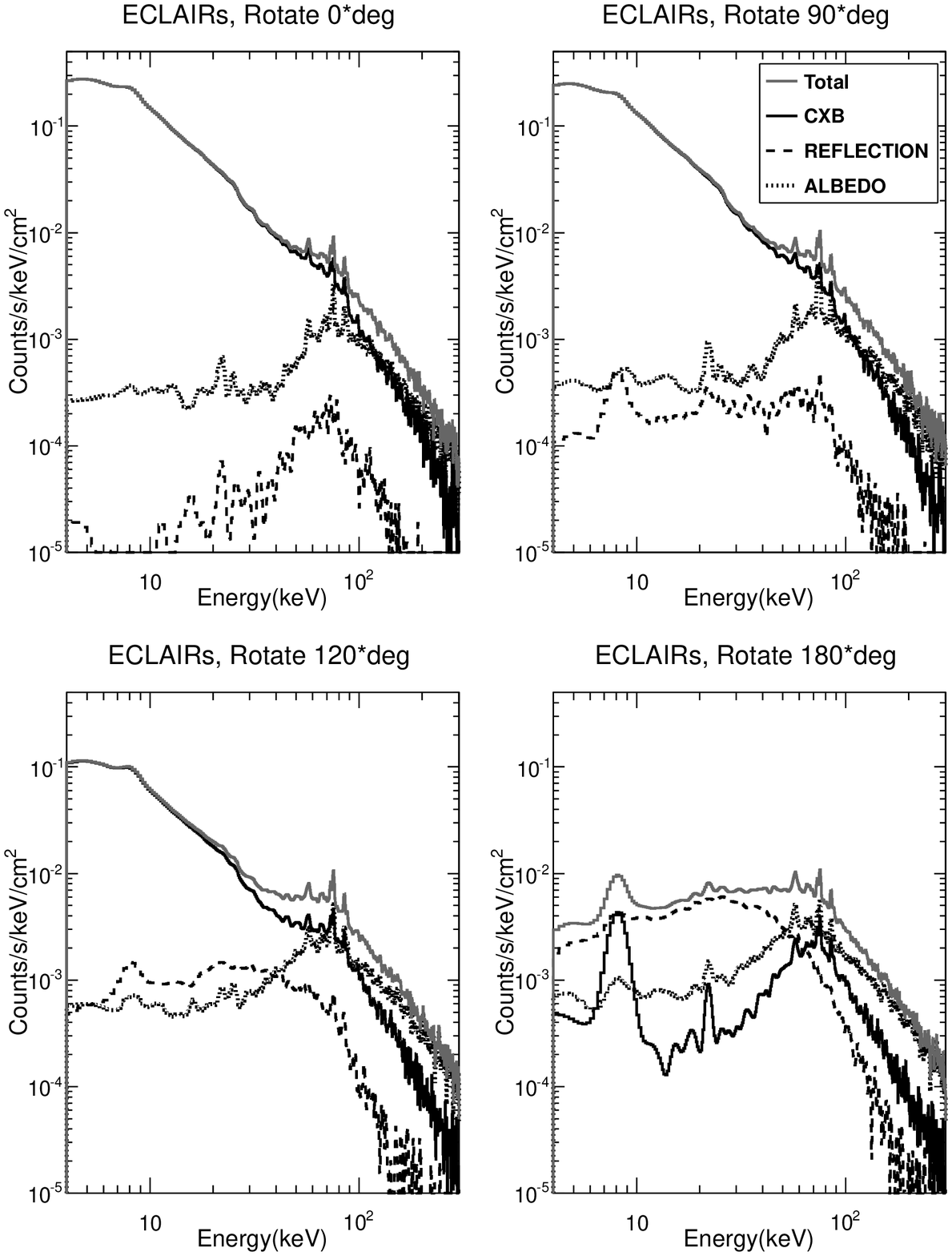}
  \caption{Three kinds of gamma-ray background of ECLAIRs change with the Earth positions.}
  \label{fig:Cxg2}
 \end{figure}

\subsection{ECLAIRs gamma-ray background}
\label{sec:ECLAIRs}
The same background models are applied to ECLAIRs and the results are shown in
Figure \ref{fig:Cxg1} and Figure \ref{fig:Cxg2}.
The results are analyzed in the same way as what has been done for GRM.

From Figure \ref{fig:Cxg1}, the CXB reaches its maximum level when the Earth is at Rotate 0$^\circ$ and 
it reaches its minimum level when the satellite faces the Earth. The ALBEDO and REFLECTION 
components increase gradually as the Earth moves close to the FoV. Figure \ref{fig:Cxg2} shows the CXB becomes 
lowest at relatively low energies when the Earth reaches Rotate 180$^\circ$. 
The ALBEDO is always the highest in the relatively high energy range. 
The REFLECTION plays a more and more important role when the Earth moves near the FoV. 

There are a few peaks in the background spectra due to element emission lines, 
including the peaks of 8 keV and 75 keV from the element Copper and Lead in the shield respectively, 
and the peak of 57 keV from the element Tantalum in the mask.
The most intense Cu-Ta-Pb lines will be used for energy calibration. 

The total background is presented in Figure \ref{fig:Cxg1} (right bottom). The Earth occultation results 
in the background with marked difference at low energy (below $\sim$50 keV).  And the highest background level 
is obtained when the Earth is at Rotate 0$^\circ$. At energies above 100 keV, 
the highest background level is obtained when the Earth is at Rotate 180$^\circ$.

%%TABLE 1
%
\begin{table*}[h]
\caption{Counting rate (counts/$cm^{2}$/s) of each background
component in various energy ranges.}
\label{tab:1}
\begin{center}
%\tiny
%\footnotesize
\scriptsize
\renewcommand{\arraystretch}{1.3}
\begin{tabular}{|c|l|l|l|l|l|l|l|l|}
\hline
			    & Earth	  	 &  			&  4-30  &  30-60       &  60-250   &  250-550     &  0.55-1 	       &  1-5 \\
\raisebox{2.2ex}[0pt]{Instrument} 
			  &  Positions  	&\raisebox{2.2ex}[0pt]{Background}	     & keV	  &  keV	      &  keV	  & keV   &  MeV    &  MeV	\\
\hline
			&			  	 & Cosmic     & -		   & 0.423		& 0.263	& 0.018		& 0.002		& 0.001 \\ \cline{3-9}
			&Rotate 0$^\circ$	 & Albedo	   & -		   & 0.005		& 0.036	& 0.029		& 0.013		& 0.015 \\ \cline{3-9}
GRM\_NaI		&			 	 & Reflection    & -		   & 0.001		& 0.003	& 0.000		& 0.000		& 0.000 \\ \cline{2-9}

			&  				& Cosmic & -	   & 0.007		& 0.023	& 0.006		& 0.001		& 0.000 \\ \cline{3-9}
			&Rotate 180$^\circ$	& Albedo   & -	   & 0.108		& 0.335	& 0.107		& 0.036		& 0.034 \\ \cline{3-9}
			&			        & Reflection & -	   & 0.204		& 0.119	& 0.001		& 0.000		& 0.000 \\ \cline{1-9}

\hline
			& 				& Cosmic  & -	    & -		& 0.184	& 0.026		& 0.005		& 0.002 \\ \cline{3-9}
			&Rotate 0$^\circ$	& Albedo    & -	    & -	   	&0.438	& 0.196		& 0.079		& 0.095  \\ \cline{3-9}
GRM\_CsI		&			    	& Reflection & -	    & -	   	& 0.063	& 0.002		& 0.000		& 0.000  \\ \cline{2-9}

			&  				 & Cosmic & -	   & 0.007		& 0.023	& 0.006		& 0.001		& 0.000 \\ \cline{3-9}
			&Rotate 180$^\circ$    & Albedo   & -	   & 0.108		& 0.335	& 0.107		& 0.036		& 0.034 \\ \cline{3-9}
			&			         & Reflection & -	   & 0.204		& 0.119	& 0.001		& 0.000		& 0.000 \\ \cline{1-9}

\hline
			&  			 	& Cosmic	& 2.460	& 0.258	& 0.180	& - 			& -		  	 & -	 \\ \cline{3-9}
			&Rotate 0$^\circ$	 & Albedo	& 0.009	& 0.018	& 0.109	& -			& -			 & -    \\ \cline{3-9}
  ECLAIRs		&			   	 & Reflection	& 0.001	& 0.002	& 0.008	& -			& -			 & -	  \\ \cline{2-9}

			&				& Cosmic	& 0.016	& 0.032	& 0.108	& -			& -			& -	  \\ \cline{3-9}
			&Rotate 180$^\circ$	& Albedo   	& 0.023	& 0.073	& 0.214	& -			& -			& -	  \\ \cline{3-9}
			&			     	 & Reflection	& 0.119	& 0.115	& 0.051	& -			& -			& -	 \\ \cline{1-9}

%%%%
\hline
\end{tabular}
    \begin{list}{}{}
      \item - Invalid energy range for the corresponding instrument.
     \end{list}
\end{center}
\end{table*}

%%TABLE 1

\subsection{Impact on the sensitivity and the GRB detection rate}
\label{sec:sensitivity}
\begin{figure}[!ht]
  \centering
  \includegraphics [scale=0.6]{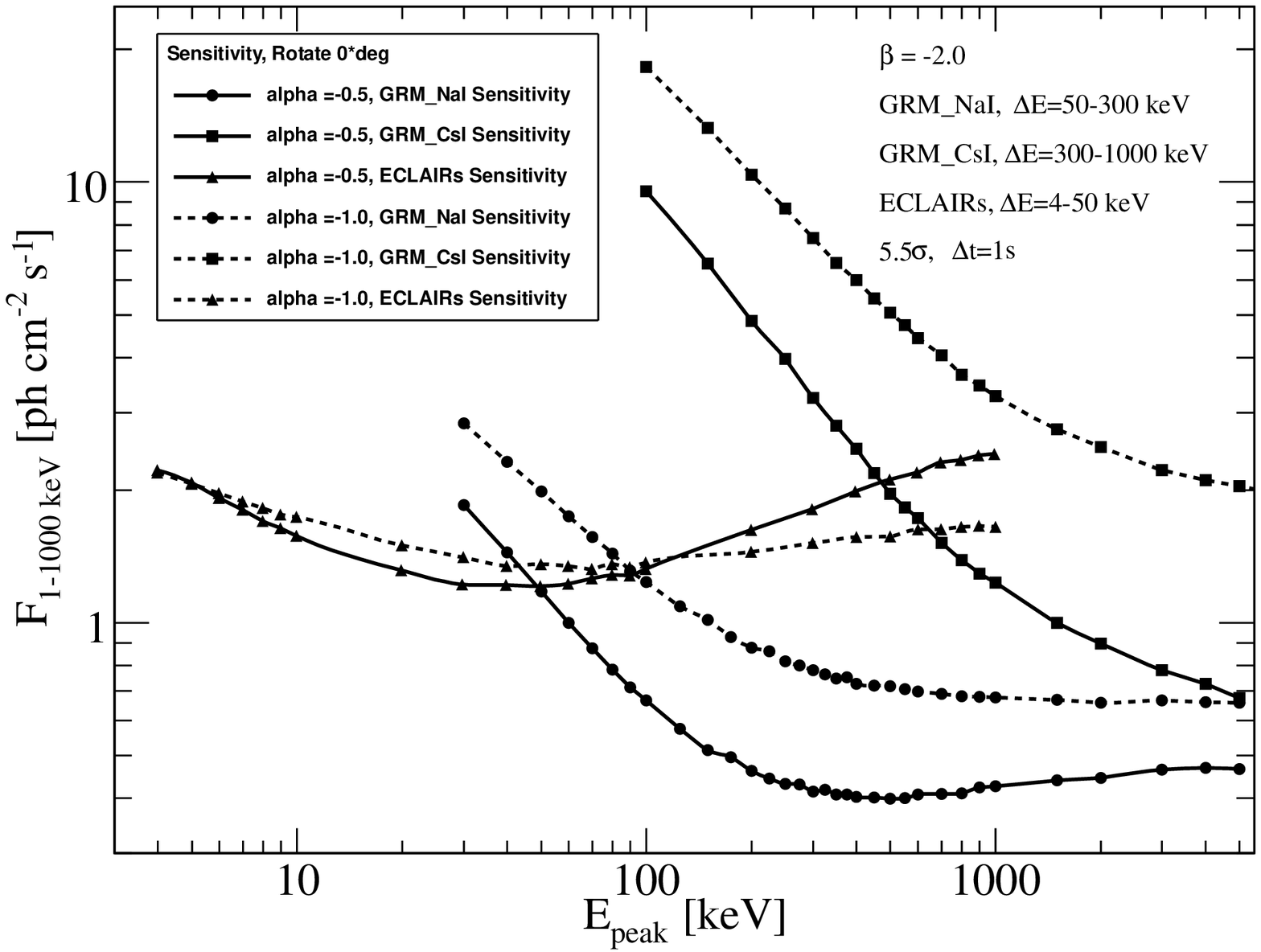}
  \caption{Threshold peak fluxes of GRM and ECLAIRs in energy range 1-1000 keV as a function 
of the GRB peak energy Epeak. Solid lines: $\alpha$ = -0.5, $\beta$= -2. Dashed lines: $\alpha$ = -1, $\beta$= -2..}
  \label{fig:Sensitivity}
 \end{figure}

The background variations for GRM and ECLAIRs with the relative Earth positions
defined in Section \ref{sec:models} are presented in Section \ref{sec:GRM} and
Section \ref{sec:ECLAIRs}. Based on these simulations, the peak fluxes at
detection threshold in the energy range 1-1000 keV were computed using the method described in \cite{Band2003}. 
In the computation, we considered two BAND spectra \cite{Band1993} of GRBs: one spectrum with $\alpha$ = -0.5, $\beta$ = -2, 
the other one with $\alpha$ = -1, $\beta$ = -2. GRBs located on the axis of the FoV, and the background level 
with Earth position Rotate 0$^\circ$ was used. The integration time was 1 s and the significance $\sigma$ was 5.5. 
The 4-50 keV and 50-300 keV energy bands are selected respectively for ECLAIRs and GRM considering their efficiency. The results are shown in Figure \ref{fig:Sensitivity}.

In general, ECLAIRs is sensitive to the GRBs with low Epeak and slightly more sensitive to the harder GRBs 
with $\alpha$ = -0.5 when Epeak $<$ 100 keV. GRM is sensitive to GRBs with relatively high Epeak, 
especially to the harder bursts with $\alpha$ = -0.5. As shown in Figure \ref{fig:Sensitivity}, comparing with ECLAIRs, 
GRM\_NaI is more sensitive for bursts with Epeak $>$ 50 keV when $\alpha$ = -0.5, 
and more sensitive for bursts with Epeak $>$ 90 keV when $\alpha$ = -1. GRM\_CsI is sensitive to 
the hard GRBs with high Epeak. The sensitivity of GRM and ECLAIRs are complementary which makes 
SVOM sensitive in a wide Epeak range, which allows accurate Epeak measurements.

%Table 2
%\multirow{nrows}[bigstructs]{width}[fixup]{text}
%\multicolumn{cols}{pos}{text}
\begin{table*}[h]
\caption{The sensitivity of GRM\_NaI $^\dagger$ and ECLAIRs $^*$.}
\label{tab:2}
\begin{center}
\scriptsize
% \tiny 
\renewcommand{\arraystretch}{1.3}
\begin{tabular}{|l| l| l| l| l| l| l| l|}
\hline	

				  & \multicolumn{3}{c|}{GRM\_NaI}					& \multicolumn{4}{c|}{ECLAIRs}		\\ \cline{2-8}
Earth				&FoV		& Background 		& Sensitivity 		& FoV	&$\overline{S}$	& Background 		& Sensitivity  \\	
Positions			&(sr)		& (counts/cm$^2$/s)	& (ph/cm$^2$/s)	& (sr)		 & (cm$^2$) 	&(counts/cm$^2$/s)	& (ph/cm$^2$/s) \\ 

\hline
Rotate 0$^\circ$		& 2.5 	& 0.41			& 0.22		& 2.02 		& 236.8 			& 1.10 		& 1.62	\\
\hline 
Rotate 30$^\circ$	& 2.5 	& 0.42			& 0.22		&  2.02 		& 236.8 			& 1.11 		& 1.63	\\
\hline
Rotate 60$^\circ$	& 2.5 	& 0.43			& 0.22		& 2.02 		& 236.8 			& 1.15 		& 1.65 	\\
\hline	
Rotate 90$^\circ$	& 2.2 	& 0.44			& 0.22		& 1.74 		& 254.3 			& 1.06 		& 1.53  	\\
\hline
Rotate 120$^\circ$	& 1.1 	& 0.44			& 0.22		& 0.91 		& 219.8 			& 0.74 		& 1.46	\\
\hline
Rotate 150$^\circ$	& 0.7 	& 0.53			& 0.24		& 0.56 		& 64.9 			& 0.57 		& 3.10	\\
\hline
Rotate 180$^\circ$	& 0.0 	& 0.61		& N/A			&0.0 			& 0.0				& 0.64 		& N/A 	\\

\hline
\end{tabular}    
\begin{list}{}{}
      \item $^\dagger$ 50-300 keV, $\sigma$=5.5, exposure time = 1 s.
      \item $^*$ 15-150 keV, $\sigma$=7.0, exposure time = 1 s.
\end{list}
\end{center}
\end{table*}

%End Table 2
%

\begin{figure}[!ht]
  \centering
  \includegraphics [scale=0.6]{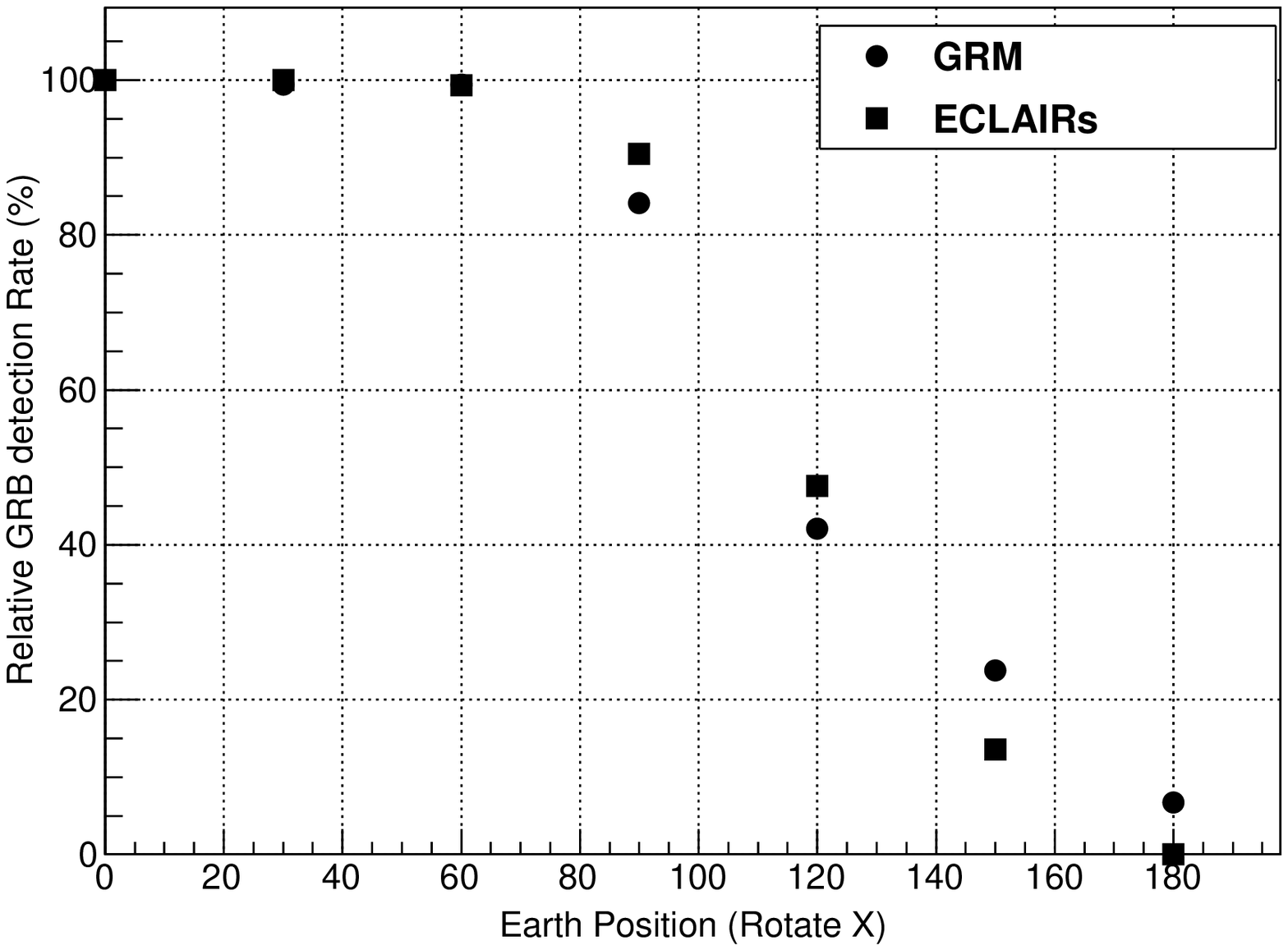}
  \caption{Evolution of the relative GRB detection rate of GRM\_NaI and ECLAIRs as a function of the Earth positions.}
  \label{fig:Ngrb}
 \end{figure}

Based on the background-simulation results in the previous section, the corresponding sensitivities (Table \ref{tab:2})
and GRB detection rates (Figure \ref{fig:Ngrb}) were calculated in the specific energy range with the given significance considering 
the GRB trigger efficiency, the observation-time loss due to SAA passages, etc. 

In Table \ref{tab:2}, the FoV is the Field of View which is not obstructed by the Earth;
the Sensitivity, which is the minimum flux that can be detected, is derived from the formula (\ref{equ:snr}).
For ECLAIRs, the Earth occultation results in the background with magnitude
difference at low energy (see Figure \ref{fig:Cxg1}).
This induces the value of sensitivity (1.46 ph/cm$^{2}$/s) corresponding to Rotate 120$^\circ$ less than 
that (1.62 ph/cm$^{2}$/s) corresponding to Rotate 0$^\circ$ (see Table
\ref{tab:2}). This means that the appropriate Earth occultation can improve the
sensitivity of ECLAIRs. Taking into account the energy range 4-50 keV, the
sensitivity can be improved by 25\% ( $\frac{Sensitivity(Rotate 0^\circ) - Sensitivity(Rotate 120^\circ )}{Sensitivity(Rotate 0^\circ)}$ ).

\begin{equation}
\label{equ:snr}
 \sigma = \frac{Sensitivity \cdot Eff}{ \sqrt{Sensitivity \cdot Eff + Background} } \sqrt{Area \cdot Time}
\end{equation}

In this work, $\sigma$=5.5 for GRM and $\sigma$=7 for ECLAIRs. 
Eff means the detection efficiency when the incident photons are perpendicular to the detector plane 
in the specific energy range which is 50-300 keV for GRM and 15-150 keV for ECLAIRs. 
The efficiency of ECLAIRs is computed taking the average GRB power-law spectrum with index of -1.64 of SWIFT as input spectrum. 
For GRM, the input spectrum is the average GRB BAND spectrum of BATSE ($\alpha$ = -1.1, $\beta$ = -2.3, Epeak = 266 keV) \cite{Kaneko2006}. 
For ECLAIRs, the Area is $\overline{S}$ in Table \ref{tab:2} which is the weighted mean area illuminated by GRBs, 
considering the closed fraction (60\%) of the coded mask, the GRB distribution in the whole FoV and the Earth occultation. 
For GRM, the Area is 567 cm$^2$ which is the whole area of the detection crystal for two GRDs. 
At variance from ECLAIRs, GRM has no capability to locate GRBs and to distinguish the energy deposit coordinate in the detector. 
The Background in the formula is in unit of counts/cm$^2$/s. 
Time in the formula is equal to 1 second for both GRM and ECLAIRs.

In the GRB detection-rate calculations, we assume that each Earth position lasts 1/7 of the whole satellite orbit period.
For every Earth position, we calculate the GRB detection rate separately, 
and sum all of them together to get the total number of GRBs detected per year.
The Earth occultation, the observation-time loss due to SAA passages and the trigger efficiency are considered. 

The GRB detection-rate calculation of ECLAIRs is based on Swift observation. 
And the factor caused by the different low-energy thresholds between ECLAIRs and Swift is also considered.
For GRM, the GRB detection-rate calculation is based on BATSE observation and uses the algorithm given by BAND \cite{Band2002} 
which was used by Dong \cite{Dong2009}. Within the assumptions, we can detecte 66 GRBs/year with ECLAIRs and
74 GRBs/year with GRM taking into account the crystal NaI only. Applying the same method to the crystal CsI, 7 GRBs/year are obtained.
We take the GRB detection rate coressponding to Rotate 0$^\circ$ as 100\%. 
And then we get the relative GRB detection rate for each Earth position as shown in Figure \ref{fig:Ngrb}. 

The main idea of the algorithm of BAND \cite{Band2002} is to compare the sensitivity to that of BATSE in the energy range 50-300 keV. 
It is a very good algorithm to estimate the GRB detection rate of GRM using the data of NaI. But for using the data of CsI, 
it is not a good choice because of its low detection efficiency ($\sim$5.5\%) in the energy range 50-300 keV. 
The sensitive energy of CsI is above 300 keV, and its detection efficiency is approximately 38\% in 300-1000 keV 
taking into account the average GRB Band Spectrum of BATSE. The number of hard GRBs detected by SVOM can 
probably be increased by finding other methods to use the data of CsI in the energy range 300-1000 keV. 
This study will be performed in a further work.

\section{Conclusion}
\label{sec:conc}
In this work, three kinds of background varying with Earth positions have been simulated for GRM and ECLAIRs which are
two high-energy instruments aboard the SVOM mission. In particularly, the simulations of ALBEDO with the model of Sazonov2007 
are described in detail for its complexity and precision. Following the background simulations, the corresponding sensitivities 
and GRB detection rates are calculated based on the given observations. 

Concerning the impact of each background component on the instruments, generally, 
for both GRM\_NaI and ECLAIRs, the cosmic gamma-rays are the most important background source 
when the Earth is outside of the FoV, especially in the relatively low energy range. 
The albedo gamma-rays are always the most important background source in the relatively high energy range. 
The reflected cosmic gamma-rays, which are lower than cosmic gamma-rays in flux and lower than albedo gamma-rays in energy, 
have the least impact on the instruments, but are still an important component in the low energy range when the Earth enters the FoV. 
For GRM\_CsI, albedo gamma-rays are always the dominant background source.

As for the spectral models used in the simulations, the limitation lays in the energy range. 
ECLAIRs is sensitive from 4 keV to 250 keV and GRM is sensitive from 30 keV to 5 MeV. 
However, the model of Moretti2009 was obtained considering the data in the energy range 1.5-200 keV. 
The model of Sazonov2007 is a function of the energy which covers from 25 keV to 300 keV. 
Only the energy range of the Churazov2006 model reaches the MeV range. As a result, one further work 
on the gamma-ray background estimations is to find better source models for the wide energy range of SVOM. 
Maybe different models should be used for different energy ranges.

In the calculations in Section \ref{sec:sensitivity}, only gamma-ray background components are considered without including other kinds of backgrounds 
induced by protons, electrons, positrons, etc. as well as the background produced by the instrument itself. 
However, we cannot conclude that the value of the GRB detection rate is overestimated for this reason.
Because there are many factors affecting the final results in fact. For example, to compute the sensitivity values 
in Table 2 accurately, the efficiency corresponding to different incident directions of photons from the FoV 
should be considered instead of only the vertical direction. Furthermore, the sensitivity used to trigger GRBs on board 
will consider diverse energy ranges, integral time scales and significance levels. One assumption in our computation is that 
the satellite has seven states relative to the Earth positions and each state lasts 1/7 of the satellite orbit period. 
This is a simplification, since the real pointing law on orbit will be more complex. 
Back to our gamma-ray background simulations, the maximum ALBEDO spectrum with the minimum solar modulation 0.25 GV 
at the highest level orbit position (30$^\circ$N, 287$^\circ$E) was used. Accordingly, a more accurate estimation of 
the GRB detection rate needs further work and another article will be prepared.

\section*{Acknowledgements}

This work was supported by 973 Program 2009CB824800, NSFC10978001 and the Knowledge Innovation Program 
of the Chinese Academy of Sciences, under Grant No. 200931111192010. Donghua ZHAO gratefully acknowledges 
China Scholarship Council and CEA/Saclay where most of this work was done.


\begin{thebibliography}{00}

 \bibitem{Paul2011} J. Paul, J. We, S. Basa, \& S.-N. Zhang,  C. R. Phys., 12 (2011) 298
 \bibitem{Basa2008} S. Basa, J. Wei et al., SF2A. Conf. (2008) 161.
 \bibitem{Ghirlanda2006} G. Ghirlanda, G. Ghisellini et al., New Journal of Physics 8 (2006) 123.
 \bibitem{Cordier2008} B. Cordier, Desclaux, F. et al., AIP Conf. Proc. 1000 (2008) 585.
 \bibitem{Dong2009} Y.-W. Dong, B.-B. Wu et al., Sci. China. Ser. G-Phys. Mech. Astron. 52 (2009) 1.
 \bibitem{Schanne2007} S. Schanne, B. Cordier et al., Proc. 30thICRC, Merida, Mexico (2007).
 \bibitem{Godet2009} O. Godet, P. Sizun et al., Nucl. Instr. Meth. A 603 (2009) 365.
 \bibitem{Mandrou2008} P. Mandrou, S. Schanne et al., AIP Conf. Proc. 1065 (2008) 338
 \bibitem{Agostinelli2003} S. Agostinelli et al., Nucl. Instr. Meth. A 506 (2003) 250.
 \bibitem{Moretti2009} A. Moretti, C. Pagani et al., Astron. Astrophys. 493 (2009) 501.
 \bibitem{Sazonov2007} S. Sazonov, E. Churazov et al., Mon. Not. R. Astron. Soc. 377, (2007) 1726.
 \bibitem{Churazov2006} E. Churazov, S. Sazonov et al., (2006) (arXiv:astro-ph/0608252)
 \bibitem{Churazov2007} E. Churazov, R. Sunyaev et al., Astron. Astrophys. 467 (2007) 529.
 \bibitem{Godet2005} O. Godet, PhD Thesis, “Monte-Carlo simulations of the wide-field coded mask camera of the GRB mission ECLAIRs”, (2005).
 \bibitem{Lg2009} G. Li, M. Wu, S. Zhang, Y. Jin, Chinese Astronomy and Astrophysics 33, (2009) 333.
 \bibitem{Ajello2008} M. Ajello, J. Greiner et al., Astrophys. J. 689 (2008) 666.
 \bibitem{Sreekumar1998} P. Sreekumar, D. L. Bertsch et al., Astrophys. J. 494 (1998) 523
 \bibitem{Gruber1999} D. E. Gruber, J. L. Matteson et al., Astrophys. J. 520(1999) 124.  
 \bibitem{Usoskin2005} I. G. Usoskin, K. Alanko-Huotari et al., J. Geophys. Res., Vol.110, A12108 (2005) 
 \bibitem{Band2003} D. L. Band , Astrophys. J. 588 (2003) 945.
 \bibitem{Band1993} D. Band, J. Matteson et al., Astrophys. J. 413 (1993) 281.
 \bibitem{Kaneko2006} Y. Kaneko, R.D. Preece et al., Astrophys. J. 166 (2006) 298.
 \bibitem{Band2002} D. L. Band, Astrophys. J. 578 (2002) 806.
 
\end{thebibliography}
\end{document}